\documentclass[nofootinbib,aps,superscriptaddress,11pt]{revtex4-1}
\usepackage{geometry}
\usepackage{graphicx}
\usepackage{dcolumn}
\usepackage{bm}
\usepackage{epsfig,amsmath}
\usepackage{comment}
\usepackage{amssymb}
\usepackage{subfigure,esint}
\usepackage{float}
\usepackage[usenames,dvipsnames]{color}
\usepackage[pagebackref=false, colorlinks=true]{hyperref}
\definecolor{redish}{rgb}{0.7,0.2,0.0}  
\definecolor{bluish}{rgb}{0.2,0.5,0.8}

\hypersetup{linkcolor=redish,          
                  citecolor=blue,        
                  filecolor=magenta,      
                 urlcolor=blue}          

\begin{document}
\author{Kunal Pal}\email{kunalpal@iitk.ac.in}
\affiliation{
Department of Physics, Indian Institute of Technology Kanpur, \\ Kanpur 208016, India}
\author{Kuntal Pal}\email{kuntal@iitk.ac.in}
\affiliation{
Department of Physics, Indian Institute of Technology Kanpur, \\ Kanpur 208016, India}
\author{Tapobrata Sarkar}\email{tapo@iitk.ac.in}
\affiliation{
Department of Physics, Indian Institute of Technology Kanpur, \\ Kanpur 208016, India}
\title{\Large Conformal Fisher information metric with torsion}

\bigskip

\begin{abstract}
We consider torsion in parameter manifolds that arises via 
conformal transformations of the  Fisher
information metric, and define it for information geometry of a wide class of physical systems. 
The torsion can be used to differentiate between probability 
distribution functions that otherwise have the same scalar curvature and hence
define similar geometries. In the context of thermodynamic geometry, our
construction gives rise to a new  scalar - the torsion scalar defined on the manifold, while retaining
known physical features related to other scalar quantities. We analyse 
this in the context of the Van der Waals and the Curie-Weiss models. In both
cases, the torsion scalar has non trivial behaviour on the spinodal curve. 

\end{abstract}
\maketitle

\section{Introduction and motivation}
\label{Intro}

The Fisher information metric (FIM) corresponding to a probability distribution function (PDF) 
is ubiquitous in the study of information geometry \cite{amari}.
Variants have been studied in physics for decades, in systems ranging from fluids,
spin chains and black holes. The literature on the topic 
is by now vast, and some representative works can be found in \cite{tapo1}-\cite{tapo5} and references therein. 
For reviews, see \cite{ruppeiner},\cite{brody}. 

Recall that a PDF $P_{\xi}(x)$ is considered to be parameterised by real parameters 
$\xi^{1},\cdots,\xi^{n} \in \Re^{n}$, and is a function of stochastic 
(or physical) variables $x$ ($ x \in \chi$, the space of the physical variables). 
It is assumed that the PDF is well behaved, i.e., 
it is a $C^{\infty}$ function, and satisfies
\begin{equation}
\label{normalisation}
P_{\xi}(x) \geq 0 ~ \forall x \in \chi~,~~\int  	P_{\xi}(x) dx =1~.
\end{equation}
Then the Fisher information metric corresponding to this PDF  is defined by  
\begin{equation}
\label{FIM_definition}
g_{ij}= E_{\xi}\big[\big(\partial_{i}l_{\xi}\big)\big(\partial_{j}l_{\xi}\big)\big]~,
\end{equation}
where  $l_{\xi}(x)=l(x;\xi)=\ln P_{\xi}(x)$ and $E_{\xi}$ refers to the expectation value of $\xi$
with weight $P_{\xi}(x)$. Also, partial derivatives are with respect to $\xi$, so that the FIM 
is defined on an $n \times n $ manifold.

Given a PDF, it is in general straightforward to define the FIM (we can write it in a closed form 
provided that the integrals are analytically tractable),
but the inverse processes, i.e., finding a PDF from a given FIM is not unique \cite{clingman}. 
The difficulty lies in the fact that there can be two  (in principle infinitely many) different 
PDFs that give same FIM. Moreover, the FIM may have extra symmetries that were not originally 
present in the PDF \cite{erdmenger1}.  
A well known illustration is provided by the Gaussian and the Cauchy distributions,  
\begin{equation}
\label{gaussian}
P_G(x; \sigma, \mu)= \frac{1}{2\pi \sqrt{\sigma}}\exp\bigg[-\frac{(x-\mu)^2}{2\sigma^{2}}\bigg]~,~~
P_C(x;x_{0},\gamma)=\frac{1}{\pi}\bigg[\frac{\gamma}{\gamma^2+(x-x_{0})^2}\bigg]~,
\end{equation}
respectively, where in the definition of  $P_G$, $ \sigma$ and $\mu$ denote respectively 
the standard deviation and the mean. The FIMs of the two metrics are similar : 
\begin{equation}\label{gaussianfisher}
ds_G^2=\frac{d\mu^2+ 2d\sigma^2}{\sigma^2}~,~~ds_C^2 = \frac{dx_{0}^2+ d\gamma^2}{2\gamma^2} ~,
\end{equation}
and both are Euclidean versions of the $AdS_2$ metric, i.e., are hyperbolic spaces with constant curvatures. 
Clearly then, the map from space of PDFs to FIM is many to one,
in principle, infinity to one. Recently, this problem has been addressed in \cite{clingman}, \cite{erdmenger1}.
One of the purposes of this paper is to offer a simple resolution by considering metrics related to
$ds_G^2$ and $ds_C^2$ via conformal transformations (CTs) and further by introducing torsion on 
the parameter manifold. We will then discuss the implications of these in fluid and spin systems. 

In this context, note that the exponential class of PDFs mentioned 
above, can in general be written as 
\begin{equation}
\label{exp}
P(x,\theta)= \exp\Big[C(x)+\theta^{i}F_{i}(x)-\Psi(\theta)\Big]~,
\end{equation}
where $C(x), F_{i}(x)$ are functions of the physical variable $x$ only, and $\Psi(\theta)$ is 
a function of $\theta$ only. Here $\theta_i$ are called the canonical coordinates for this distribution, and  
$\Psi(\theta)$ is called the potential, which  is  also the normalisation factor of the PDF.
For example, for the Gaussian distribution, we can identify the potential function to be 
\begin{equation}
\label{pai_gaussian_2}
\Psi_G(\mu,\sigma)= \frac{\mu^2}{2\sigma^2}+ \ln(\sqrt{2\pi}\sigma)~.
\end{equation}  

We are interested in the exponential distribution for two main reasons : firstly, the calculation 
of the  FIM in this case becomes simple, since it depends only on the second derivatives
of the potential function, taken with respect to the canonical coordinates, i.e.,
$g_{ij}=\partial_{i}\partial_{j}\Psi(\theta)$.
Secondly, if we consider a physical system in its thermodynamic limit (where the particle number 
$N \rightarrow \infty$) then the PDF can be thought to satisfy
a Boltzmann distribution at inverse temperature 
$\beta =\frac{1}{T}$, namely, $P(x,\theta)= Z^{-1}e^{-\beta H(x, \theta)}$,
where $H(x, \theta)$ is the Hamiltonian of the system, and $Z$ is the partition function. 
This PDF thus belongs to the
 exponential family : $P(x,\theta)=\exp[\beta H(x,\theta)-\ln Z(\theta)]$. 
Here $\theta_i$ represent coordinates of the parameter space, which can be coupling constants of the theory, 
or can be the thermodynamic variables.
Since the partition function of a thermodynamic system  
is $\ln Z = -\beta F$, with $F$ being the free energy, then defining the reduced free energy
per site  $f=-\beta F/N$, one obtains the metric on the parameter manifold to be 
$g_{ij}= \partial_{i}\partial_{j}f (\theta)$.

In the context of thermodynamic systems, scalar quantities such as the Ricci scalar $R$ and the
expansion scalar $\Theta$ of a geodesic congruence constructed out of the information metric play a crucial role. 
Several properties of such scalars are well known in two dimensional systems, for example,\\
{\bf A}. Generally, $R$ diverges along the spinodal curve. \\
{\bf B}. The sign of $R$ captures the nature of interactions in a system (attractive or repulsive). \\
{\bf C}. The universal scaling relation $R\sim \lambda^{-2}$ near criticality, where $\lambda$ is an affine
parameter measuring geodesic length.\\
{\bf D}. The universal scaling relation $\Theta \sim \lambda^{-1}$ near criticality. \\
{\bf E}. The relation $R \sim l^d$ near criticality, where $l$ is the correlation length.\\
{\bf F}. The equality of $R$ at first order phase transitions (which is an alternative 
characterisation of such transitions and distinct from, say, the Maxwell construction in Van der Waals
fluids).\\
One can reasonably demand that any modification to the information metric should keep the 
features {\bf A} to {\bf D} intact, as these follow purely from geometry. For example, a 
CT that depends on the PDF would produce different (distinguishable) metrics starting 
from a possibly degenerate set, but simply doing this would cause undesirable changes in $R$ 
(see eq. (\ref{Rchange}) below)
that might, for example, violate these features. A similar situation occurs with $\Theta$, as we explain
via eq. (\ref{Cth}). Importantly, for relations {\bf E} and {\bf F} to be physically justifiable, $R$
should have dimensions of volume, due to which thermodynamic geometry is always defined via potentials
per unit volume (for a more detailed discussion in the context of black holes, see \cite{tapo6},\cite{tapo7}). 

Keeping these issues in mind, in this paper we shall introduce a new metric on the parameter 
manifold which is conformally related to the known FIM, but in which $R$     
is invariant under the CT and so is $\Theta$, so that properties {\bf A} to {\bf D} are retained. 
As explained below, for this to be true, we have to
drop the assumption of symmetric connection on the parameter manifold i.e., we have to consider
torsion as well as curvature on the manifold. If we  assume a particular form of transformation 
of the connection and torsion components under a CT, then it can be shown that the curvature tensor 
and the curvature scalar are related by a benign algebraic factor (that does not contribute to scaling relations) 
under a CT, thereby retaining the usefulness of the original FIM. 
With the help of the (different) torsions \textit{induced} on the parameter manifold, we can also predict the difference between the 
two different PDF, that cannot  otherwise be done with the help of same FIM. 
As we shall see, we  consider the transformation of torsion under a CT in such a way that the 
transformation rule that keeps the curvature tensor invariant also maps the geodesic of 
one metric the geodesic of transformed one so that the expansion scalar is invariant under 
the process, which can be used to retain property {\bf D} along with {\bf A} to {\bf C} above. 
With these motivations, we move on to present our main arguments. 


\section{Fisher information metric with torsion}

We  drop the assumption 
of symmetric connection on the parameter manifold (but keep the metricity requirement intact), 
i.e., we consider a parameter manifold with torsion \cite{nakahara}. However, instead of providing a predetermined 
prescription for inducing torsion on the manifold, we shall resort to a 
CT (with the conformal factor depending on the PDF) to change (or generate) 
the torsion. Given two PDFs (call them $P_{1}$ and $P_{2}$) we will associate each with a 
conformal factor (say $\omega_{1}^2$ and $\omega_{2}^2$), which depends on the nature of the PDFs under 
consideration. Then instead of calculating the FIMs (call them $g^{(1)}_{ab}$ and $g^{(2)}_{ab}$ respectively) 
corresponding to these PDFs, we will calculate  metrics related to the FIMs by CTs. 
The resulting metrics, which we will call conformal Fisher information metric (CFIM), 
$\bar{g}^{(1)}_{ab}$ and $\bar{g}^{(2)}_{ab}$ (denoted by a bar) are naturally  different from corresponding FIMs. 
Since $\omega_{1}^2$ and $\omega_{2}^2$ are different then $\bar{g}^{(1)}_{ab}$ and 
$\bar{g}^{(2)}_{ab}$ are also different from each other, even if 
$g^{(1)}_{ab}$ and $g^{(2)}_{ab}$ are same. As argued subsequently in section \ref{cotorsion}, the presence of a 
non-symmetric connection and hence the torsion is crucial here. Even if we want the Ricci 
scalars $\bar{R}^{1}$ and $\bar{R}^{2}$ calculated form the barred metrics to predict different 
physical properties, we also want the scalars $R^{1}$ and $\bar{R}^{1}$ to predict same behaviour, 
simply because they represent same physical system,   
and one way to make this possible is to introduce torsion on the parameter manifold.

Before concluding this section we consider the following important question.
As we have seen   above, given a PDF we can compute the related FIM by using the standard
formula of Eq. (\ref{FIM_definition}). If  the FIM corresponding to a given PDF is related
to the CFIM by a CT, then what is the relation between the PDF which has  CFIM as the FIM,
with the PDF of the original FIM? In appendix - \ref{disformal} we show that if two PDFs
are related by a function of coordinates on the parameter manifold, then the corresponding
FIMs are related by a disformal transformation, which is a generalization of the CT. 
The properties of a metric related to the FIM by a disformal 
transformation  will be pursed further in a separate work.  

\section{Conformally related Fisher information metrics}

\subsection{Conformal transformation with  symmetric connection is not enough}

As we have discussed, as a means of circumventing the problem of 
degenerate FIM corresponding to two widely different PDFs we shall use a metric related 
to the FIM by a CT.  Specifically, we consider two FIMs related by a conformal transformation of the form 
\begin{equation}\label{conformal}
\bar{g}_{ab}=\omega^{2}(\xi)g_{ab}~, ~~~\text{where}~~~ 
\omega^{2}(\xi)=\exp\big({2\Omega (\xi)}\big)~,
\end{equation}
with  $\xi$ being the coordinates on the parameter manifold and $\omega^{2}(\xi)$ 
is generic real positive function of the coordinates, and see how the physically 
meaningful quantities constructed from them are related to one another.
For the present discussion, the exact form of the conformal factor $\omega^2 $ is not important, 
instead, our main aim is to point out that  conformally transformed metric with  
symmetric  (Levi-Civita) connection  on the manifold  does not capture the same physics 
as that of the ordinary FIM. 

To see this, note that when the usual symmetric Levi-Civita  connection is used, 
the Ricci scalar of a $n$ dimensional manifold   transform under the CT
in the following way (see, e.g., \cite{scarroll},\cite{robwald}) 
\begin{equation}
\label{Rchange}
R \rightarrow \bar{R} =  e^{-2\Omega}\Big(R-2(n-1)\Box \Omega- (n-2)(n-1)(\nabla\Omega)^{2} \Big)~.
\end{equation} 
Here $\Box$ denotes the Lapalcian  operator with respect to the metric $g_{ab}$.
Thus the Ricci scalar is not invariant under CT, even when $n=2$ for which the second term above is zero. 
\footnote{By saying that a quantity, say 
$\Phi$ (which can be be either a  field or a geometric quantity such as Ricci scalar), is 
conformally invariant we mean that there exists a real number  $c$ such that,  
as $g_{ab} \rightarrow \bar{g}_{ab}=\omega^{2}g_{ab}$ it transformed 
to $\Phi \rightarrow \bar{\Phi}=\omega^{c}\Phi$  \cite{robwald}.\label{note5}}
If we consider the metric $g_{ab}$
as the FIM of a physical  system then the Ricci scalar $R$ is the meaningful quantity 
that can be constructed from it. If we use $\bar{R}$ instead of $R$,
it is clear from the above expression that the conclusions about the system (as elaborated upon in
the introduction) based on $\bar{R}$ would be different from ones based on $R$.
This is something we need to avoid, and will thus need to make the Ricci scalar of the transformed metric invariant
under a CT.

We can understand the problem with the CT with symmetric connection 
from a different  point of view as well, namely by considering geodesics in the parameter manifold. 
Let us consider a geodesic $\theta^{a} (\tau)$ on the parameter manifold, $\tau$ being
the affine parameter along the geodesic. We denote the tangent vector  along the geodesic 
as $u^{a}=\partial \theta^{a}/\partial \tau$ and which  is assumed to be normalized as $u^{a}u_{a}=1$. 

Using the usual transformation formula for the symmetric connection under a CT,
 we can easily calculate the relation between the acceleration vectors before and
 after the transformation as
\begin{equation}\label{acc}
\bar{u^{a}}\bar{\nabla}_{a}\bar{u}_{b}= u^{a}\nabla_{a}u_{b}+\big(g_{ab}
+u_{a}u_{b}\big)\nabla^{a}\Omega~.
\end{equation}
We now see that the geodesics of the metric $\bar{g}_{ab}$ (which are essentially the solutions 
of $\bar{u^{a}}\bar{\nabla}_{a}\bar{u}_{b}=0$) are not geodesics of the metric $g_{ab}$ (which satisfy $u^{a}\nabla_{a}u_{b}=0$). 
The force free motion in terms of one metric become the force equation 
in another, the force being proportional to the derivative the conformal factor and directed 
perpendicular to the velocity. 

Now if we consider a collection of  non intersecting geodesics (a geodesic congruence) on 
the parameter manifold we can have another scalar quantity characterising these geodesics, 
namely the expansion scalar $\Theta$ of the congruence. 
This expansion scalar is just the trace of the extrinsic curvature of a hypersurface on the manifold, 
and  in terms of the velocity vector it is given by the formula $\Theta=\nabla_{a}u^{a}$. 
It was argued in \cite{kumarsarkar}, in the context of thermodynamic geometry, that
close to the critical point this scalar also diverges, having a particular power 
law dependence with the affine parameter. 
This power law dependence is different from that of the Ricci scalar, and also characterises 
the phase transitions of a physical system in the thermodynamic limit. 

It can be also checked that when the connection is symmetric, the expansion scalar is also not 
invariant under a CT. Rather it changes as 
\begin{equation}
\label{Cth}
\bar{\Theta}=e^{-\Omega}\big(\Theta+(n-1)u^{a}\nabla_{a}\Omega\big)~.
\end{equation}
In passing, we note that the change of the expansion scalar along the geodesic equation is encoded 
in the Raychaudhuri equation. Taking a derivative of both sides of Eq. (\ref{Cth}) with respect to 
the respective affine parameter, one can check that the Raychaudhuri equation is also not conformally invariant. 

These then are some of the problems associated with a CFIM. We will now discuss how to
circumvent these issues.

\subsection{Conformal transformation with non-symmetric connection}\label{cotorsion}

Keeping the above discussions in mind, we want  the transformed FIM to keep the curvature 
scalar (as well as  expansion scalar) invariant under CT. 
It is known that this can be achieved using a non-symmetric connection.\footnote{This is  known in the context of 
gravitational physics and cosmology (see, e.g., \cite{lucat}), 
but to the best of our knowledge this has not been used in the context of information geometry.} 
The non-symmetric connection naturally introduces another structure - torsion on the parameter 
manifold \cite{nakahara}. If we demand the torsion tensor to transform in a particular way under a CT
 then it can be shown that the geodesic equation and the expansion 
scalar are invariant (see below). It is important to keep in mind that  the metric and the torsion 
on a manifold are two independent constructions and the CT of the 
metric does not specify the transformation of torsion under the CT. 
This is also true for a non-symmetric  connection (as opposed to the Levi-Civita 
connection which transforms under CT in a fixed way). 
This freedom of making the transformation of the torsion and the connection is 
what is responsible for the invariance of the geodesics. Let us now see how this comes about.
In what follows, $(x,y)$ and $[x,y]$ will denote the symmetric and antisymmetric combinations
of two quantities $x$ and $y$, respectively. 

The general connection on a manifold satisfying the metricity 
condition can be written as \cite{nakahara, frenkel, shapiro,  poplawski}
\begin{equation}
{\Gamma}^{a}_{b c} = {{\Gamma}^\circ}^{a}_{b c} + \Big(T^{a}_{*bc}+ T_{bc *}^{\hspace{4mm}a} 
+ T_{c b *}^{\hspace{4mm}a}\Big)~,
\end{equation}
where the $\circ$ over the first term refers to the  Christoffel symbols which can be obtained 
form the metric and is symmetric in the lower indices, and the stars indicate the position of 
the upper index. The group of terms in the brackets is called the \textit{contorsion}.
Now, consider the CT of the form of Eq. (\ref{conformal}), under which 
the  Christoffel symbols change by
\begin{equation}
\delta{{\Gamma^\circ}}^{a}_{bc} = \delta^{a}_{b}\nabla_{c}\Omega +
\delta^{a}_{c}\nabla_{b}\Omega -
g_{bc}\nabla^{a}\Omega~.
\end{equation}

On the other hand the CT of the metric does not specify the transformation 
of the torsion, and in principle we could choose not to transform it at all, in fact there are 
different way torsion can change under a CT (see for instance \cite{shapiro}). 
For our present purpose, let us consider the following transformation rule for the connection 
: $\bar{{\Gamma}}^{a}_{bc}= {\Gamma}^{a}_{bc}+\delta{\Gamma}^{a}_{b\gamma}~,$
i.e. we assume that the total connection transforms linearly under the CT. 
Then we specify the following transformation rule for the contorsion
\begin{equation}\label{Cotra}
\delta K^{a}_{\hspace{4mm}bc} = g_{bc}\nabla^{a}\Omega-\delta^{a}_{c}\nabla_{b}\Omega~.
\end{equation}
To see how this comes about consider the transformation for the first term in the  contorsion 
\begin{equation}\label{torchange1}
\delta T^{a}_{*bc} =  \delta^{a}_{[b}\nabla_{c]}\Omega= \frac{1}{2}\Big(\delta^{a}_{b}\nabla_{c}\Omega-\delta^{a}_{c}\nabla_{b}\Omega\Big)~.
\end{equation}
Similarly, for the other two terms, we have  respectively,
\begin{equation}\label{torchange23}
\delta T_{bc *}^{\hspace{4mm}a} =\frac{1}{2}\Big(g_{bc}\nabla^{a}\Omega- 
\delta^{a}_{b}\nabla_{c}\Omega \Big)~,~~\text{and}~~
  ~~~ \delta T_{cb *}^{\hspace{4mm}a} = \frac{1}{2}\Big(g_{bc}\nabla^{a}\Omega- 
  \delta^{a}_{c}\nabla_{b}\Omega \Big)~.
\end{equation}
Adding the last two transformations we get \cite{shapiro, lucat} 
\begin{equation}\label{T2nd}
2\delta T_{(bc) *}^{\hspace{4mm}a} = g_{bc}\nabla^{a}\Omega -\delta^{a}_{(b}\nabla_{c)}\Omega~.
\end{equation}
The transformation rules of Eq. (\ref{torchange1}) and Eq. (\ref{T2nd}) are sufficient to 
satisfy the contorsion transformation rule of Eq. (\ref{Cotra}).  
It is important to notice  how the particular symmetries of the torsion 
tensor (anti-symmetric in last two indices) and contorsion tensor (anti-symmetric in first 
two indices) are manifested in the above transformation rules.

With the transformation rule for each components of the connection, the change in connection can be written simply
as: $\delta{\Gamma}^{a}_{bc} = \delta^{a}_{b}\nabla_{c}\Omega~.$
Now, it is a  straightforward to see that  the curvature tensor, defined in terms of the 
connection as $R^{a}_{\hspace{2mm}bcd}= \nabla_{c}\Gamma^{a}_{bd} -\nabla_{d}\Gamma^{a}_{bc} 
+ \Gamma^{a}_{e c}\Gamma^{e}_{bd} -\Gamma^{a}_{e d}\Gamma^{e}_{bc}~,$
is invariant under this transformation  i.e. 
$ R^{a}_{\hspace{2mm}bcd}  \rightarrow \bar{R}^{a}_{\hspace{2mm}bcd}$.
Similarly the contraction of curvature tensor (analog of the Ricci tensor) is also left invariant 
$\bar{R}_{ab} = R_{ab}~.$
Finally,  the transformed  curvature scalar curvature is related to the previous one by a 
simple overall conformal factor,
\begin{equation}
\label{R_with_torsion}
R=g^{ab}R_{ab} ~~~\rightarrow ~~~ \bar{R}=\bar{g}^{ab}\bar{R}_{ab} 
= \omega^{-2} g^{ab}R_{ab}=\omega^{-2}R~.
\end{equation}
Thus when non-symmetric connections are used, the Ricci scalar is invariant under
the CT (we again emphasise that by ``invariant,'' we mean related by a multiplicative factor which does not affect
scaling relations). 

The implication of the above observations for our purpose  is that, the conclusions about 
a statistical system based on the scalar curvature of the corresponding FIM will remain 
unchanged even after the CT of the FIM, provided that we assume the 
presence of a non-symmetric connection. Obviously, even though the corresponding scalar curvatures 
are related by an overall factor, the conformally transformed FIM is not same as 
the FIM. However, the key point is that this new metric captures the same information as that of 
the FIM, as detailed in points {\bf A} to {\bf D} in the introduction. 

Since the conformal factor is smooth and is a positive function of the coordinates, the sign and the
nature of divergence of the scalar curvature of the FIM remain intact (here we are assuming that
the CT is not a singular CT). 
The other important 
point to note is that, though the scalar curvature remains invariant, 
for two different conformal factors chosen to distinguish the same 
FIM originating from two different PDFs, the corresponding torsions \textit{induced} on 
the parameter manifold are not same, i.e., depend on the conformal factors, which in turn 
depend on the PDFs we are considering and gives a useful \textit{measure} that can distinguish 
the two PDFs  - that cannot otherwise be done with the FIM only. Here it is important 
to note that the torsion is induced on the parameter manifold by means of the 
CT in a pre-determined way. 

We now turn to the expansion scalar. We start by noticing that when the connection
changes as $\delta{\Gamma}^{a}_{bc} = \delta^{a}_{b}\nabla_{c}\Omega~,$  the covariant derivative of the
velocity vector $u^{a}$ changes as
\begin{equation}\label{velocitytor}
\bar{\nabla}_{a}\bar{u}^{b}= \big(\partial_{a}\bar{u}^{b}
+\bar{\Gamma}^{b}_{\hspace{2mm}ca}\bar{u}^{c}\big)
=e^{-\Omega}\big(\partial_{a}u_{b}+\Gamma^{b}_{\hspace{2mm}ca}u^{c}\big)
=e^{-\Omega}\big(\nabla_{a}u_{b}\big)~. 
\end{equation}
Taking the trace of both sides we see that the expansion scalar is conformal invariant
as well $\Theta~ \rightarrow ~\bar{\Theta}= e^{-\Omega}\Theta$. Similarly 
one can also show that the geodesics equation and the Raychaudhuri equation are invariant. 
Thus after the CT  both the scalar curvatures show the same divergent 
behaviour near criticality as that of the original  one.

\subsection{Example : Separating the Gaussian and Cauchy distributions}

Till now, the discussion has been general, and applicable to any well defined conformal transformation. 
Now we shall specify the conformal factor. In this paper we propose to
take the exponential of the normalisation factor $N(\xi)$ of a  PDF in as the 
conformal factor connecting between the FIM and the CFIM. 
The normalisation factor is a function of the  coordinates on the parameter manifold and
characterises the PDF uniquely, hence this choice is one of the simplest choice one can make.  
Note that now with a non-symmetric connection, we can immediate use the last relation of
Eq. (\ref{R_with_torsion}) to distinguish between the Ricci scalars of two PDFs that were otherwise identical. 

In this section we shall illustrate the procedure by considering  the two PDFs 
in Section \ref{Intro}, namely the Gaussian and Cauchy distributions of eq. (\ref{gaussian})),  
which have the same scale invariant hyperbolic metric as the FIM. 
To distinguish these two PDFs in terms of metrics on the parameter
manifold, we perform a CT with the conformal factors taken to be 
the  exponential of the normalization factors $N_{i}$ of the $i$th PDF ($i=1,2$) :
\begin{equation}
\omega_{i}^{2}(\xi) =\exp\big({2\Omega_{i}(\xi)}\big)
=\exp\big({N_{i}(\xi)}/N_0\big) ~.
\end{equation}
Here $N_0$ is a  constant  having same dimension as $N_i$s. In the expressions below
we shall omit this for brevity, and it is to be understood that in order to make the analysis
dimensionally consistent, we suitably set a dimensionfull quantity to unity.  

For the Gaussian distribution, the potential function $\Psi(\mu,\sigma)$ is the
normalisation factor \cite{erdmenger1}. So the transformed version of the first relation in
Eq. (\ref{gaussianfisher}), with $N_1 = \Psi(\theta_1,\theta_2)$ ($\theta_{1},\theta_{2}$ are the 
canonical coordinates), is given by 
\begin{equation}\label{conformalgaussian}
g^{(1)}_{ab} \rightarrow \bar{g}^{(1)}_{ab}= \exp \big({2\Omega_{1}(\xi)}\big) g^{(1)}_{ab}= \exp \big({\Psi(\theta_1,\theta_2)}\big)g^{(1)}_{ab}~.
\end{equation}
And similarly the transformed version of the second relation in Eq. (\ref{gaussianfisher}), with $N_2 = \pi/\gamma$, can be written as 
\begin{equation}\label{conformalcauchy}
g^{(2)}_{ab} \rightarrow \bar{g}^{(2)}_{ab}= \exp \big({2\Omega_{2}(\xi)}\big) g^{(2)}_{ab} 
= \exp (\pi/\gamma) g^{(2)}_{ab}~.
\end{equation}
As can be seen the barred metrics are different from each other, even if the unbarred 
versions are of the  same form in the respective coordinates. Hence using the last relation in Eq. (\ref{R_with_torsion}),
the new Ricci scalar can be used to distinguish between them. 
Before moving on  here we note  that we consider the CT as an exact change of geometry,
\footnote{This is not a proper set of diffeomorphisms, which imply simultaneous
coordinate rescaling as well as the Weyl rescaling of the fields, see
\cite{quiros}.} such that two metrics ($g_{ab}$ and $\bar{g}_{ab}$) characterize two different 
geometries on the same statistical  manifold, and  they are written in terms of  the same coordinate charts.

We now compute torsions on the respective manifolds which are   
`generated' by means of the CT. If we assume that the initial 
parameter manifold is torsion free, then from Eqs.  (\ref{conformalgaussian}) and (\ref{torchange1}) 
the torsion induced on the 
parameter manifold of Gaussian distribution is given by \footnote{This assumption can be easily dropped, 
and we are only specifying the change of torsion by Eqs.
(\ref{torchange1}) and  (\ref{torchange23}). However, general calculation with non-zero torsion is 
straight forward but lengthy.}
\begin{equation}\label{T_components}
T^{a}_{*b c}= \frac{1}{2}\delta^{a}_{[b}\partial_{c]}\Psi~,
\end{equation}
where, as above we have used, $\Omega= \Psi(\theta_1,\theta_2)/2$. Since  the 
parameter manifold is two dimensional with coordinates $(\theta_1,\theta_2)$,  only 
four  components of torsion tensor are non-zero.  
Furthermore, in general, in two dimensions, the  torsion 
has only two independent components - known as  the torsion traces. Explicitly, these can be written as
\begin{equation}
T^{1}_{*21}=-\frac{1}{4}\frac{\partial\Psi (\theta_{1},\theta_{2})}
{\partial \theta_{2}}=\frac{2\theta_{2}-\theta_{1}^2}{16\theta_{2}^2}=-\frac{1}{4}\big(\mu^2+
\sigma^2\big)~,~~
T^{2}_{*12}=-\frac{1}{4}\frac{\partial\Psi (\theta_{1},\theta_{2})}
{\partial \theta_{1}}=\frac{\theta_{1}}{8 \theta_{2}}=-\frac{1}{4}\mu~,
\end{equation}
where the final expressions are written in terms of the $(\mu, \sigma)$ coordinates.

On the other hand, the torsion components induced on the parameter manifold of the 
Cauchy distribution are given by
\begin{equation}
T^{1}_{*21} =\frac{\pi}{4\gamma ^2}~, \hspace{4mm}     T^{2}_{*12}= 0~ .
\end{equation}
Here only one component is non zero. 
Clearly, if we describe the parameter manifolds of the two PDFs with  coordinate 
charts  so that two manifolds have same metric induced 
on them, then they have different induced torsions through the CT, and hence by this torsion 
we can distinguish between them. 

Of course, one needs a scalar quantity to make any concrete statement about such
a distinction. To this end, we define the torsion scalar \footnote{This is the formula for the 
torsion scalar for a two dimensional parameter manifold. For higher dimensional manifolds this formula will be 
generalized \cite{AG}.}
 $\mathcal{T}=T_{abc}T^{abc}$, from which
it can be checked that 
\begin{equation}
\mathcal{T}_{Gaussian} = -\frac{e^{-\frac{\mu ^2}{ 2\sigma ^2}} \left(\mu^2+
\sigma ^2\right)^2}{16 \sqrt{2\pi} \sigma ^5}~,~~
\mathcal{T}_{Cauchy} = \Big(\frac{\pi }{2\gamma  }\Big)^2 e ^{- \frac{ \pi}{\gamma}} ~.
\end{equation}
Like the transformed Ricci scalar, the torsion scalar also distinguished between the two PDFs. 

\section{Torsion in fluid and spin systems}

Now it is natural to ask how a non-symmetric connection affects the information geometry of
a statistical system in the thermodynamic limit (or analogously, the Ruppeiner's thermodynamic geometry \cite{ruppeiner}), where
a potential is known and the formalism above can be applied conveniently. 
To this end, we will briefly discuss the effects of such a connection on a few statistical
systems that show phase transitions. As we have noticed before, in the thermodynamic limit, 
the probability distribution functions of these systems belong to the exponential class,
and hence the intensive potential is the normalization factor of these distribution. 
These potentials can be obtained by calculating the partition functions \cite{brody}. For these cases, 
our proposal is to take $\omega^2(\theta)=\exp (f(\theta))$, where $f(\theta)$
is the intensive thermodynamic potential, as the   conformal factor connecting the FIM and CFIM. 
The components of the torsion tensor can then be obtained from Eq. (\ref{T_components}) with $\Psi$
replaced by $f(\theta)$, while the torsion scalar is given by $\mathcal{T}=T_{abc}T^{abc}$.
 
Our first example is the Van der Waals (VdW) fluid system, where we use the
metric corresponding to Ruppeiner's geometry in the usual thermodynamic coordinates, whose expression is 
already known \cite{ruppeiner}. Specifically,
one takes the thermodynamic potential as the free energy per unit volume, given by \cite{LandauLifshitz}
\begin{equation}
f_{{\rm VdW}} = -\rho T{\rm ln}\left(\frac{e}{\rho}\right) + \rho c_v T{\rm ln}\left(\frac{e}{T}\right) - 
\rho T{\rm ln}\left(1 - b\rho\right) - a\rho^2~,
\end{equation}
where $a$, $b$ are the coefficients appearing in the VdW equation of state denoting its departure from the ideal gas, 
$c_v$ is the specific heat at constant volume, and $\rho=1/v$ and $T$ are the number density per molecule of fluid and 
the temperature, respectively. We work with a reduced equation of state in terms of $\rho_r = \rho/\rho_c$, $T_r = T/T_c$ and 
set $\rho_c = T_c = 1$. The information metric (with $c_v = 3/2$) is given here by 
\begin{equation}
ds^2 = \frac{3}{2} \frac{\rho_r}{T_r^2}dT^2 + \frac{9\left[4T_r - \rho_r\left(\rho_r - 3\right)^2\right]}
{4\rho_rT_r\left(\rho_r - 3\right)^2}d\rho^2~.
\end{equation}
With this choice of the conformal factor and in coordinates $(\theta^1,\theta^2)=(T,\rho)$, 
the components of the torsion tensor take very simple
forms : $T^1_{~21}$ and $T^2_{~12}$ are the chemical potential and the entropy per unit volume, respectively.
Then, following the procedure outlined above, we find that the torsion scalar $\mathcal{T}$ diverges
all along the spinodal curve, and near criticality, it behaves as $\mathcal{T}\sim t^{-1}$, where
$t = (1-T_r)$. 
\begin{figure*}[h!]
\begin{minipage}[b]{0.45\linewidth}
\centering
\includegraphics[width=0.9\textwidth]{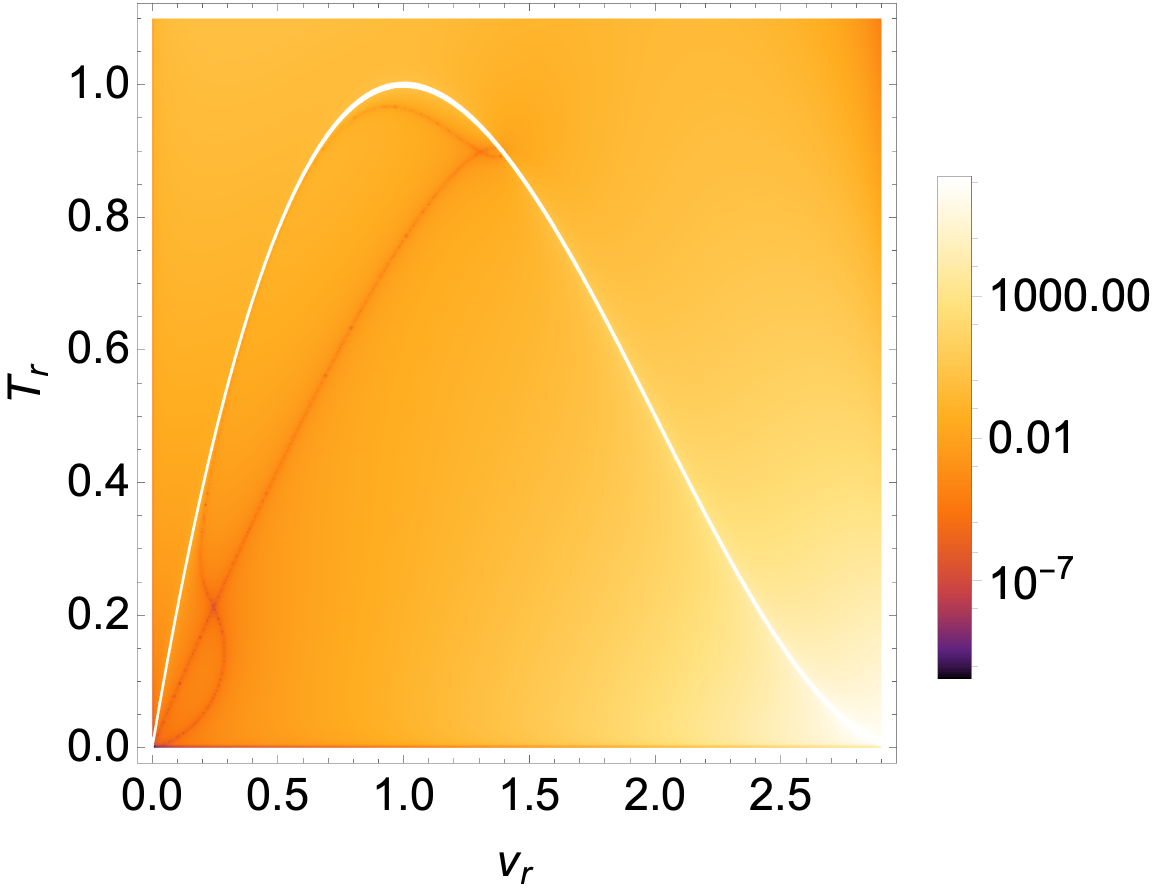}
\caption{$\mathcal{T}$ as a function of $v_r$ and $T_r$ for the VdW model, on a logarithmic scale. 
The divergence of ${\mathcal T}$ locates the spinodal curve.}
\label{torsionVdW}
\end{minipage}
\hspace{0.05cm}
\begin{minipage}[b]{0.45\linewidth}
\centering
\includegraphics[width=0.9\textwidth]{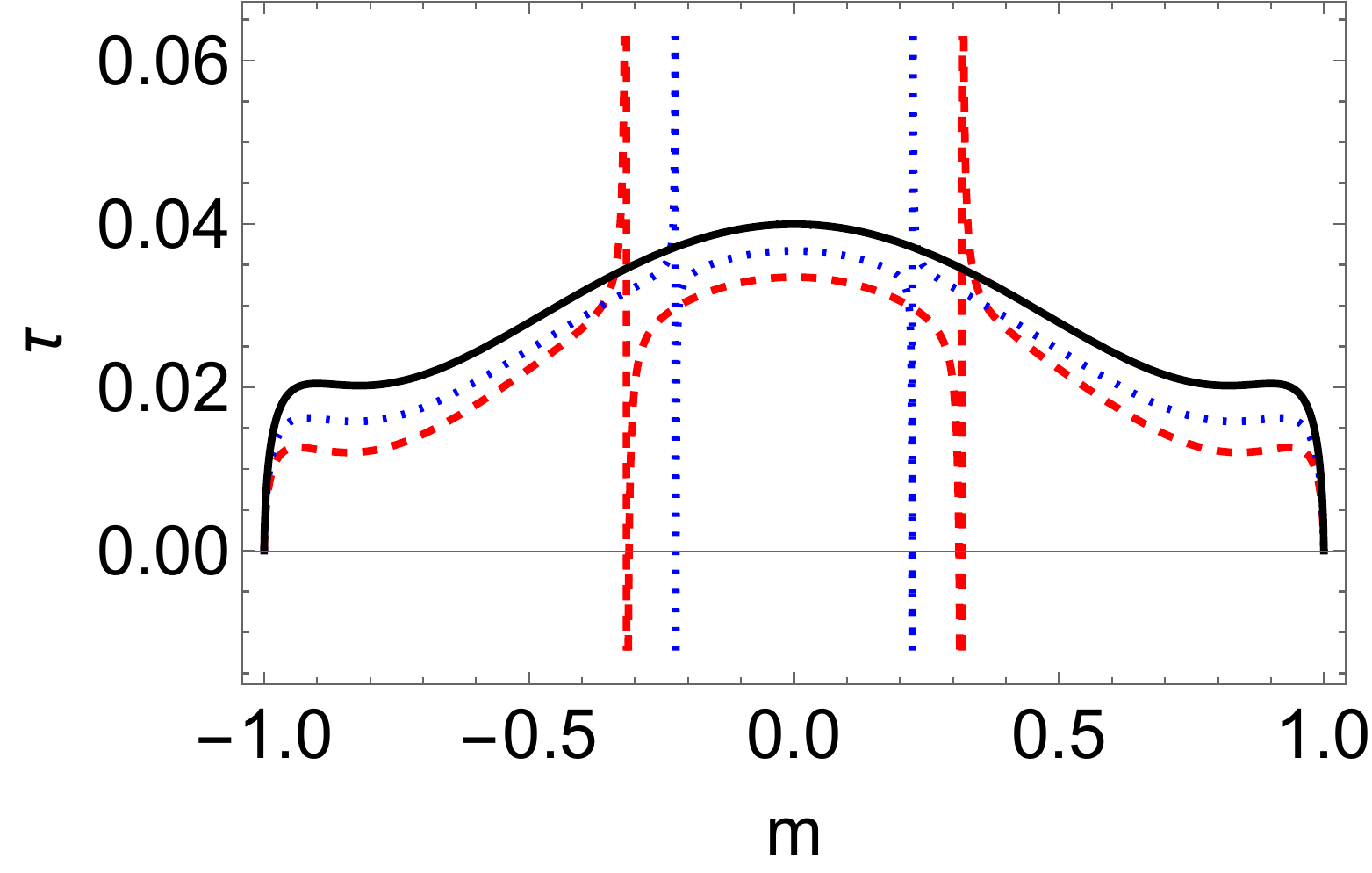}
\caption{$\mathcal{T}$ as a function of $m$ for the Curie-Weiss model, with $T_r=0.9$ (dashed red), 
$0.95$ (dotted blue) and $0.99$ (solid black).}
\label{torsionCW}
\end{minipage}
\end{figure*}
The behaviour of $\mathcal{T}$ is shown in Fig. \ref{torsionVdW} on a logarithmic scale, 
as a function of $v_r = 1/\rho_r$ and $T_r$. The divergence of $\mathcal{T}$  along the spinodal curve can be clearly seen.

Next, we consider the Curie-Weiss (CW) ferromagnetic model, which is a symmetric cousin of the VdW model, 
whose information theoretic geometry was studied in \cite{JM},\cite{tapo2}. For this model, the Helmholtz free energy 
per unit spin is given by
\begin{equation}
f_{{\rm CW}} = -k_BT{\rm ln}2 - \frac{1}{2}k_BT_cm^2 + \frac{k_BT}{2}{\rm ln}\left(1 - m^2\right) + k_BTm{\rm tanh}^{-1}m~,
\label{cwhelm}
\end{equation}
where, as before, $T$ is the temperature, $T_c$ its critical value, which we will set to unity. 
Also, $\left(\frac{\partial f_{{\rm CW}}}{\partial m}\right)_T=H$ gives the applied 
magnetic field, with $m$ being the magnetization per spin. 
It was shown in \cite{JM} that the information metric is given by  (setting $k_B=1$)
\begin{equation}
ds^2 = \frac{C_L}{T^2}dT^2 + \frac{1}{T}\frac{\left(T_c\left(1-m^2\right) -T\right)}{m^2-1} dm^2~,
\label{linecw}
\end{equation}
where $C_L$ is the ``lattice specific heat,'' an unknown function of the temperature, which arises here due to an added
mechanical energy term in the CW Hamiltonian. In fact, without this somewhat ad hoc addition, information geometry becomes 
trivial in the CW model (see  \cite{JM}). As in \cite{tapo2}, we use a convenient choice $C_L = 1 + T + T^2$. 
Our observation here is that the torsion scalar $\mathcal{T}$ is non-analytic everywhere on
the spinodal curve of this model defined as $T=1-m^2$, but becomes regular at criticality near $m\to 0$, i.e., its scaling 
exponent is zero. This behaviour is straightforward to establish analytically, and is also borne out in Fig. \ref{torsionCW}
where the dashed red, dotted blue and the solid black lines correspond to $T_r = 0.9,0.95$ and $0.99$, respectively, and the non-analyticity of
the first two and the regularity of the third on the spinodal curve are apparent from this figure. 
Finally, we have also considered the Ising model in a transverse magnetic field \cite{tapo2}
and find that the torsion scalar $\mathcal{T}$ diverges exponentially near the critical point. The computations
here are similar to the other cases, and we will skip them for brevity, and will come back to this example
briefly in the concluding section. 


\section{Conclusions and discussion}
The tools of information geometry have been used extensively to gain insights about 
coarse-grained structure of physical systems, ranging from fluids and spin chains to 
black holes. However, ambiguities can occur in this description  due to the fact that
the FIM, the primary structure used in information geometrical techniques, is a local quantity.  
Among different limitations, some well known ones include the emergence of the same FIM for different PDFs, 
as well the existence of different connections of the parameter manifold ($1$-connections, $0$-connections etc) with 
their corresponding Ricci scalars not always capturing the essential information
about the statistical system.

As a resolution to the problem of degenerate FIM corresponding to different PDFs, 
in this paper we have introduced a new quantity on
the parameter manifold, namely the torsion, by means of a CT of the FIM. 
Using the fact that the relevant scalar quantities on the parameter manifold (for example, the curvature scalar and the expansion scalar of a geodesic congruence) constructed from a metric related to the original FIM by a 
CT are conformally invariant due to presence of the torsion, we argue that scalar quantities constructed from the
conformally related FIM carry the same physical information  about 
the systems (via relations {\bf A} to {\bf D}
listed in the introduction) as that of the ones constructed from the original FIM.
 However, due to different torsions on the 
parameter manifold, different PDFs can be distinguished with our construction. Furthermore, for a few statistical 
systems whose information geometric descriptions are well known, we have 
studied the nature of the torsion scalar - a new scalar quantity defined on the manifold which is
independent of the FIM.
While this shows a divergence all along the spinodal curve for Van der Waals fluids and has a well defined
scaling relation near criticality, these features are absent for the Curie-Weiss model. 

There are two caveats in our analysis which deserve further attention. First, we have taken the
conformal factor to be the exponential of the thermodynamic potential, as a simple choice. Clearly, a multiplicative
factor with the potential might change the results, since the potential often contains explicit logarithms. This happens, for example,
in the one dimensional Ising model in a transverse field \cite{JM}. It can be checked that in this
case, if, for example, we take the conformal factor to be the exponential of twice the thermodynamic potential per particle, the
divergence of the torsion tensor vanishes at the critical point. Such an issue is not present in the
VdW and the CW models that we have studied. Secondly, as we pointed out in the introduction, in thermodynamic
geometry of fluids, one demands that $R$ has the dimensions of volume. In this context, the relevance of $\mathcal{T}$ which
naively has dimensions of energy per unit volume (upon suitably restoring factors of the Boltzmann constant)
is less obvious and needs to be studied further. 

Finally, we point out that, here we have always assumed that the metricity condition is satisfied on the parameter
manifold. The extension of our method for connections with a non-metric structure as well as with torsion is left for a future work.

\begin{appendix}
\section{Relation between FIMs for given  relation between the PDFs  } \label{disformal}
In this appendix we  derive  a relation between two FIMs corresponding to 
two given PDFs which are related by a function of coordinates on the parameter
manifold.  Let us consider two PDFs $P^1_{\xi}(x)$ and $P^2_{\xi}(x)$ which are 
described by same set of parameters $\xi$ and stochastic variables $x$, and 
assume that they are related by the following relation
\begin{equation}
	P^2_{\xi}(x)= C(\xi)P^1_{\xi}(x)~. 
\end{equation}
Here $C(\xi)$ is solely a function of the coordinates  $\xi$ of the parameter manifold. 
We want to find out the general relation between  the FIMs constructed from them, 
and since the FIM carries the information contained in a  given PDF it will 
help us to understand the nature of the PDFs as well.
	
The FIM corresponding to  $P^2_{\xi}(x)$  is given by $g^2_{ab}= 
E_{\xi}[\big(\partial_{a}l^2_{\xi}\big)\big(\partial_{b}l^2_{\xi}\big)]$ 
and  is related to the FIM corresponding to  $P^1_{\xi}(x)$ by the relation
\begin{equation}
	\begin{split}
	g^2_{ab}= C(\xi) ~ \int ~ dx ~ P^1_{\xi}(x) \Big[\partial_{a}\ln(C (\xi)P^1_{\xi}(x)) \Big]
	\Big[\partial_{b}\ln(C (\xi)P^1_{\xi}(x))\Big]\\
	= ~~C\int dx P^1_{\xi}(x) \Big[\partial_{a}\ln(P^1_{\xi}(x))\Big]
	\Big[ \partial_{b}\ln(P^1_{\xi}(x))\Big]  
	+ C\int dx P^1_{\xi}(x) \partial_{a}\ln(C) \partial_{b}\ln(C) \\
	+C\int dx P^1_{\xi}(x) \partial_{a}\ln(C) \partial_{b}\ln(P^1_{\xi}(x)) +
	C\int dx P^1_{\xi}(x) \partial_{a}\ln(P^1_{\xi}(x)) \partial_{b}\ln(C) ~.
	\end{split}
\end{equation}
	
Now, using the fact that since the PDF is normalized (see 
Eq. (\ref{normalisation})), its partial derivative is zero, 
i.e $\partial_{a}\int P^1_{\xi}(x) dx =0$, 
the last two terms of the above expression are zero, and the first two terms can 
be simplified to give the following simplified relation between two FIMs
\begin{equation}
	g^2_{ab}(\xi)=C(\xi)~g^1_{ab}(\xi)+\frac{1}{C(\xi)}\partial_{a}C(\xi) ~ \partial_{b}C(\xi)~.
\end{equation} 
	
Thus if we consider two PDFs which are related to one another by a multiplicative function
$C(\xi^{1},\cdots,\xi^{n})$ defined  on the parameter manifold, 
then the two FIMs constructed from them are related  by a general \textit{disformal} 
transformation (DT). DTs are generalization of ordinary CTs and was 
originally considered by Bekenstein \cite{JBe}. In DT two metrics are related to each 
other not only by a CT, but their  is also an extra piece which depends on the derivative of
a scalar field, which is $C(\xi)$ here.  In this paper  we have mainly focused on the FIMs related
by a CT, however it will be interesting to consider FIMs related by a DT and see if the 
scalar quantities computed from them can be made disformal invariant.

\end{appendix}

\end{document}